\documentclass[preprint,12pt]{elsarticle}

\usepackage{amssymb}

\journal{Results in Physics}

\begin{document}

\begin{frontmatter}

\title{Hydrogenation of Nd-Fe-B magnet powder under a high pressure of hydrogen}

\author{Y. Kataoka$^1$}
\author{Y. Kawamoto$^1$}
\author{T. Ono$^2$}
\author{M. Tsubota$^2$}
\author{J. Kitagawa$^1$}
\ead{j-kitagawa@fit.ac.jp}

\address{$^1$ Department of Electrical Engineering, Faculty of Engineering, Fukuoka Institute of Technology, 3-30-1 Wajiro-higashi, Higashi-ku, Fukuoka 811-0295, Japan}
\address{$^2$ Physonit Inc., 6-10 Minami-Horikawa, Kaita Aki, Hiroshima 736-0044, Japan}

\begin{abstract}
The hydrogenation of Nd$_{2}$Fe$_{14}$B under a high pressure of hydrogen has been investigated for the first time.
At the heat-treatment temperature of 600$^{\circ}$C, the almost complete decomposition of Nd$_{2}$Fe$_{14}$B into NdH$_{2+x}$ and $\alpha$-Fe is observed, although a rather long heat-treatment time is necessary to achieve the sufficient hydrogenation.
The desorption of hydrogen from NdH$_{2+x}$ does not occur in the furnace-cooling process.

\end{abstract}

\begin{keyword}
hydrogenation, Nd-Fe-B magnet, HDDR process
\end{keyword}

\end{frontmatter}

\clearpage

\section{Introduction}
Toward the improved magnetic-properties of the Nd-Fe-B permanent magnet, one of important technologies is the hydrogenation of Nd$_{2}$Fe$_{14}$B, which is employed in the so-called HDDR (hydrogenation, disproportionation, desorption, recombination) process\cite{Nakayama:JAP1991}.
The HDDR-processed magnet possesses a submicron grain-size, contributing a higher coercivity.
H$_{2}$ atmosphere with rather low pressure ($\sim$ 1 atm) is introduced in the hydrogenation process.
After the heat-treatment at 750 $\sim$ 900 $^{\circ}$C for 30 min $\sim$ 3 h, the magnet disproportionates into Nd hydride (NdH$_{2+x}$), $\alpha$-Fe and Fe$_{2}$B.

Several groups have reported the reactive milling of Nd-Fe-B powder under hydrogen at room temperature\cite{Khelifati:JMMM2000,Gang:JMMM2006,H. Lianxi:JMR2008}.
The mechanical activation of Nd-Fe-B powder leads to the decomposion into NdH$_{2+x}$ and $\alpha$-Fe, although we need high pressure of H$_{2}$ gas ($\sim$ 1 MPa).
These studies motivated us to investigate the conventional hydrogenation of Nd-Fe-B under a high pressure of H$_{2}$ gas, which has not been reported.
In this study, we have studied the hydrogenation of Nd-Fe-B under a high pressure of hydrogen.

\section{Experimental method}
We used a commercial Nd-Fe-B magnet.
The ground magnet powder, after demagnetization, weighting approximately 0.45 g were placed in a home-made cell with the volume of about 7.5 cc.
After evacuating the cell, we introduced H$_{2}$ gas of 0.45 MPa.
The cell was heated to the temperature ranging from 500 $^{\circ}$C to 600$^{\circ}$C, taking 2 h, by an electric furnace, and kept at that temperature for 6 to 84 h, followed by a furnace-cooling.
The heat-treated samples were evaluated using the powder X-ray diffraction (XRD) pattern (Cu-K$\alpha$ radiation).
We also hydrogenated Nd-Fe-B powder in a continuous H$_{2}$ gas flow at 600$^{\circ}$C for 12 h, in which the heating and cooling rates are the same as those in the above-mentioned hydrogenation process.

\section{Results and discussion}

\begin{figure}[hbtp]
\begin{center}
\includegraphics[width=0.9\linewidth]{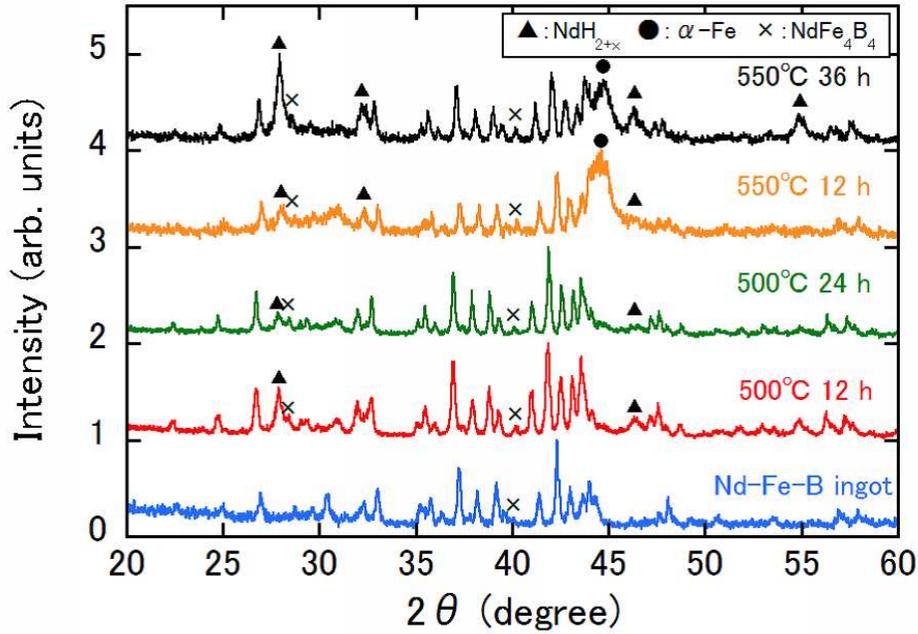}
\caption{XRD patterns of samples after heat-treatment at 500 or 550$^{\circ}$C . The XRD pattern of Nd-Fe-B ingot is also shown.}
\end{center}
\end{figure}

The XRD patterns of samples heat-treated at 500$^{\circ}$C for 12 and 24 h are shown in Fig.\ 1, where the pattern of Nd-Fe-B ingot is also exhibited.
The pattern of Nd-Fe-B ingot is mainly composed of Nd$_{2}$Fe$_{14}$B in addition to the minor phase of NdFe$_{4}$B$_{4}$.
The peaks of Nd$_{2}$Fe$_{14}$B in the samples heat-treated at 500$^{\circ}$C are shifted to lower 2$\theta$ values compared to those of the starting ingot.
This implies the lattice expansion of Nd$_{2}$Fe$_{14}$B due to the insertion of hydrogen\cite{Isnard:JAP1995}.
The decomposition of Nd$_{2}$Fe$_{14}$B does not occur because of no appearance of $\alpha$-Fe, which is confirmed by further extending the heat-treatment time to 84 h.
The NdH$_{2+x}$ phase denoted by triangles would be generated from a Nd-rich phase\cite{Rodriguez:MSE2000}.

\begin{figure}[hbtp]
\begin{center}
\includegraphics[width=0.9\linewidth]{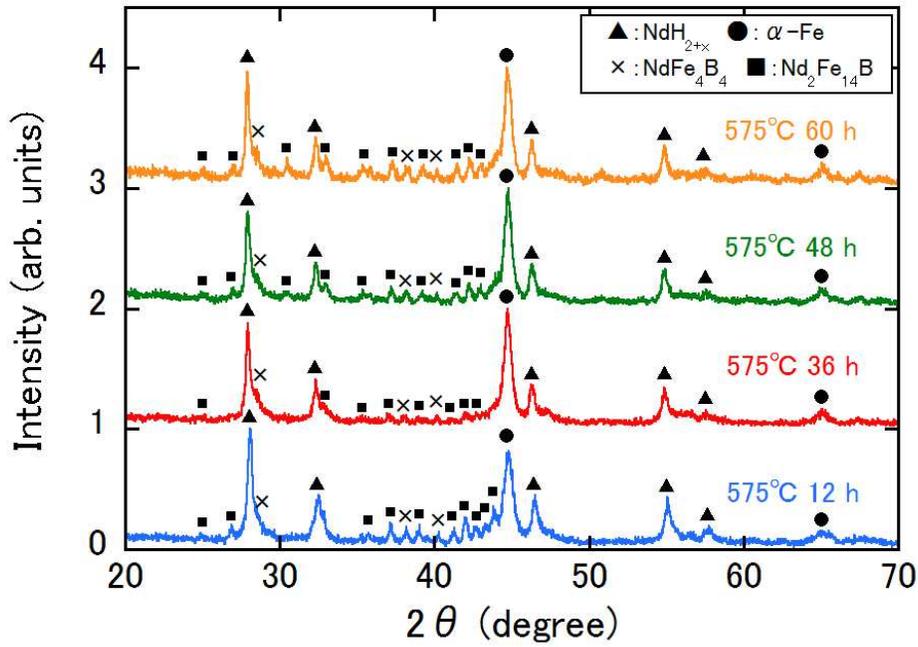}
\caption{XRD patterns of samples after heat-treatment at 575$^{\circ}$C .}
\end{center}
\end{figure}

Fig.\ 1 also shows the XRD patterns of samples heat-treated at 550$^{\circ}$C for 12 and 36 h.
The shift of peaks of Nd$_{2}$Fe$_{14}$B does not occur in each sample, indicating no insertion of hydrogen in Nd$_{2}$Fe$_{14}$B.
Although Nd$_{2}$Fe$_{14}$B is not completely decomposed, the simultaneous appearance of NdH$_{2+x}$ and $\alpha$-Fe means the partial hydrogenation of Nd$_{2}$Fe$_{14}$B.

As shown in Fig.\ 2, the hydrogenation of Nd$_{2}$Fe$_{14}$B proceeds by increasing the heat-treatment temperature to 575$^{\circ}$C.
The peaks of NdH$_{2+x}$ and $\alpha$-Fe phases dominate over those of unreacted Nd$_{2}$Fe$_{14}$B and NdFe$_{4}$B$_{4}$, which is observed throughout the heat-treatment time of interest.

\begin{figure}[hbtp]
\begin{center}
\includegraphics[width=0.9\linewidth]{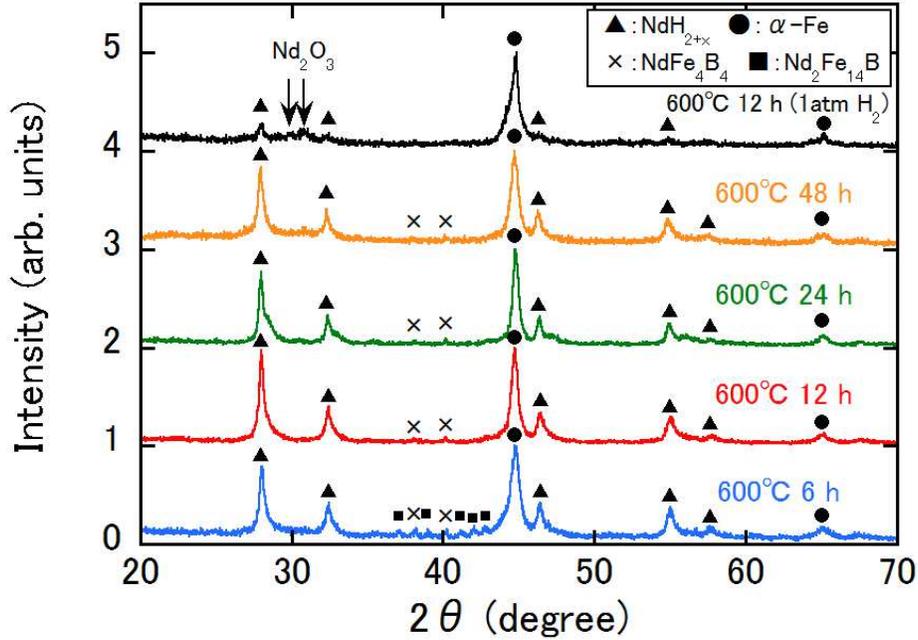}
\caption{XRD patterns of samples after heat-treatment at 600$^{\circ}$C .}
\end{center}
\end{figure}

At the heat-treatment temperature of 600$^{\circ}$C for longer than 12 h, the almost complete decomposition of Nd$_{2}$Fe$_{14}$B is observed (see Fig.\ 3).
We cannot say whether the Fe$_{2}$B phase, which is detected after the disproportionation in the HDDR process, exists or not.
To investigate the effect of high-pressure H$_{2}$ gas, we have carried out the hydrogenation of Nd-Fe-B in a continuous H$_{2}$ gas flow.
The XRD pattern of the sample is shown at the top of Fig.\ 3.
We have found that, even at the heat-treatment temperature lower than that in typical HDDR process, the complete decomposition of Nd$_{2}$Fe$_{14}$B into NdH$_{2+x}$ and $\alpha$-Fe can be achieved for the prolonged heat-treatment time.
However the peak intensity of NdH$_{2+x}$ is very weak compared to that in samples hydrogenated under high H$_{2}$-pressure.
In addition, Nd$_{2}$O$_{3}$ phase is detected (see arrows in Fig.\ 3).
During the furnace-cooling, some hydrogen might desorb from NdH$_{2+x}$ transforming into Nd$_{2}$O$_{3}$.
On the other hand, the samples hydrogenated under high H$_{2}$-pressure do not show the desorption of hydrogen after the furnace-cooling.

\section{Summary}
We have investigated the hydrogenation of Nd-Fe-B powder under the high pressure of hydrogen.
The hydrogenation of Nd$_{2}$Fe$_{14}$B proceeds almost completely at the heat-treatment temperature of 600$^{\circ}$C, although the prolonged heat-treatment time is necessary for the sufficient decomposition of Nd$_{2}$Fe$_{14}$B.

\end{document}